\documentclass[amsmath,amssymb,superscriptaddress,twocolumn,prl,showpacs]{revtex4}

\usepackage[usenames]{color}
\usepackage{graphicx}

\begin{document}

\title{Ultracold three-body collisions near overlapping Feshbach resonances}
\author{J. P. D'Incao}
\affiliation{JILA, University of Colorado and NIST, Boulder, Colorado 80309-0440, USA}
\affiliation{Institut f\"ur Quantenoptik und Quanteninformation, 
\"Osterreichische Akademie der Wissenschaften, 
6020 Innsbruck, Austria}
\affiliation{Department of Physics, Kansas State University,
Manhattan, Kansas 66506, USA} 

\author{B. D. Esry}
\affiliation{Department of Physics, Kansas State University,
Manhattan, Kansas 66506, USA} 

\begin{abstract}
We present a comprehensive collection of ultracold three-body collisions properties near
overlapping Feshbach resonances.  Our results incorporate 
variations of all scattering lengths
and demonstrate novel collisional behavior, 
such as atom-molecule interference effects.
Taking advantage of the unique ways in which
these collisions reflect Efimov physics, new
pathways to control atomic and
molecular losses open up.  Further, we show that 
overlapping resonances can greatly improve the chances of observing
multiple Efimov features in an ultracold quantum gas for nearly any system.
\end{abstract} 
\pacs{34.50.-s,67.85.-d,21.45.-v,31.15.xj}
\maketitle

In the past few years, the experimental realization of ultracold gases with multiple
species has made a whole new range of novel quantum phenomena experimentally accessible.
In many cases, the underlying many-body physics depends on both intra- and inter-species correlations,
emphasizing the importance of developing control over these types of interactions {simultaneously}. 
This control can be accomplished by applying an external 
magnetic field near values where two or more of the interactions support a Feshbach 
resonance \cite{Overlaps}, allowing for the various intra- and inter-species $s$-wave scattering 
lengths to vary drastically. Although such overlapping resonances can occur naturally, 
recent proposals \cite{ScatControl} open up the possibility of controlling the interactions
{independently} thus allowing the exploration of the multiple species strongly interacting regime. 
In this context, understanding the underlying few-body phenomena 
can potentially deepen the understanding of the many-body physics \cite{3spinsThe} as well as help to control
the much more complex collisional behavior of the ultracold gas. In fact, recently, the first experimental 
studies on the collisional behavior of ultracold gases near overlapping Feshbach resonances \cite{3spinsExp} 
have revealed interesting universal few-body physics and triggered a number of theoretical studies 
\cite{3spinsEfimov}.

The core of the few-body phenomena near overlapping Feshbach resonances is the
extension of the three-body Efimov physics \cite{Efimov} to multiple large 
scattering lengths.
Efimov physics and its manifestations in ultracold scattering 
near an isolated resonance is today well understood \cite{Us,Braaten}. 
When the scattering length $a$ 
for identical bosons, for instance, 
is tuned to the universal regime $|a|\gg r_{0}$, where $r_{0}$ is the 
range of the two-body interactions, loss rates 
reveal signatures of Efimov
states as seen in the fast-growing number of experiments \cite{EfimovExp,4bodyExp},
demonstrating the importance 
of universal few-body physics \cite{4bodyThe}.

In this Letter, we will show 
that when multiple scattering lengths are allowed to vary, the signatures 
of Efimov physics in scattering observables are much richer than the single scattering length cases 
and open up new fronts for the control of
atomic and molecular losses. 
We have found, for instance, that when two scattering lengths are large and positive, Efimov physics can cause
destructive interference in atom-molecule
collisions, allowing for longer 
molecular lifetimes, even in the absence of Pauli blocking \cite{Pauli}. 
In other cases we show that both destructive and constructive interference
can be observed in the {\em same} scattering process unlike the single resonance case. 
Importantly for few-body physics, we show that near overlapping resonances the strength of 
the effective three-body interaction behind the Efimov effect can be made much stronger and thus 
improve the observability of the Efimov effect in a great variety of systems. 
In fact, it turns out that the Efimov effect is easier to observe in {\em any}
system with overlapping resonances
than in identical boson systems. We also
derive the scattering length dependence for the important
inelastic scattering
processes in systems consisting of two identical bosons and a dissimilar particle, say $BBX$,
and three dissimilar particles, say $XYZ$, 
with two and three resonant scattering lengths. The latter is relevant 
for the recent experiments with $^6$Li in three different spin states~\cite{3spinsExp}.

\begin{figure}[htbp]
\includegraphics[width=3.1in,angle=0,clip=true]{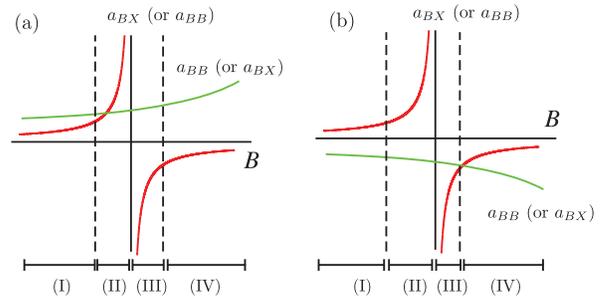}
\caption{(color online). 
Idealized picture for two overlapping $s$-wave Feshbach resonances
between  $B$ and $X$ atomic species, focusing on just one of the resonances.
In (a) there are one or two weakly bound molecular states, and in (b)
zero or one molecular state can exist. The regions (I)--(IV)
in each case exhaust all scenarios in general.}
\label{Fig1}
\end{figure}

We start our study of three-body physics near overlapping Feshbach resonances by
identifying the most generic cases. In Fig. \ref{Fig1}
we show these cases for two nearby resonances between  $B$ and $X$ atomic species, 
focusing only on the region around one of the resonances.
Figure~\ref{Fig1}(a) illustrates the case where one of the scattering lengths goes through
a pole while the other remains large and positive, while 
in Fig.~\ref{Fig1}(b) the pole occurs when the other scattering length is large and negative.
The main difference between these two cases is that in Fig.~\ref{Fig1}(a) there
exist two weakly bound molecular states $BB^*$ and $BX^*$, associated with the scattering lengths 
$a_{BB}$ and $a_{BX}$, respectively, while in Fig.~\ref{Fig1}(b) only one
weakly bound $BX^*$ molecular state
exists. Even though the cases shown in Fig.~\ref{Fig1} are only an idealization of the overlap between
two resonances, the four regions labeled in each case ---where our results and analysis are applied separately--- 
exhaust all possibilities for two overlapping resonances.

Our results for the three-body inelastic rates are derived from the generalization of our methodology \cite{Us} 
to multiple 
scattering lengths. 
The key ingredient in our approach
is that the scattering length dependence of the rates 
is determined by a combination of tunneling through 
effective three-body potential barriers
and interference effects \cite{Us}.
The form of the effective potentials thus
controls the collisional 
properties of the system.
In the adiabatic hyperspherical representation, the effective 
three-body potentials can be parameterized as
\begin{eqnarray}
W_{\nu}(R) = -\frac{s_{0}^2+1/4}{2\mu R^2} \mbox{~~~and~~~}
W_{\nu}(R) = \frac{p_{0}^2-1/4}{2\mu R^2},
\label{EffPot}
\end{eqnarray}
\noindent
for attractive and repulsive Efimov potentials, respectively, in 
each region $r_{0}$$\ll$$R$$\ll$$|a'|$ and $|a'|$$\ll$$ R$$\ll$$ |a|$, assuming $|a|$$\gg$$|a'|$.
Here, $R$ is the hyperradius describing the overall size of the system, 
$\mu$
is the three-body reduced mass \cite{Us} 
and $\nu$ the set of quantum numbers necessary to label each channel. 
The strengths of the attractive and repulsive interactions 
$s_{0}$ and $p_{0}$, respectively, as well as the systems in which they occur, 
are dictated by Efimov physics 
\cite{Braaten,Us} through the values of the mass ratios between the particles and 
whether they are of bosonic or fermionic character.
We note that $s_0$ and $p_0$ also depend on the number of resonant interactions
and will thus be different in each $R$ region.
For instance, for three identical bosons ---where all the pairs interact resonantly--- 
the attractive potential in Eq.~(\ref{EffPot}) 
has $s_{0}\approx1.00624$. On the other hand, for systems with three bosons in two different 
spin states and resonant interspecies interaction 
---giving two resonantly interacting pairs--- 
$s_{0}\approx0.41370$. Now, if the intraspecies interaction in the above example is also
resonant, the value for $s_{0}$ is the same as for three identical bosons. 
The same values are also obtained for particles in three different spin states
relevant to Refs.~\cite{3spinsExp,3spinsThe}.

\begin{table*}[htbp]
\caption{Scattering length dependence for three-body collision rates in $BBX$ and $XYZ$ systems. For $BBX$ systems
both $a_{BX}$ and $a_{BB}$ scattering lengths are resonant, while for $XYZ$ systems only $a_{XY}$
and $a_{XZ}$ are resonant. The notation $|a|$ indicates $a$$<$0, and no 
entry indicates that the associated process is not possible. Expressions for $M$ and $P$ 
are given in Eqs.~(\ref{M1})--(\ref{P2}).
  \label{TabRates}} 
\begin{ruledtabular}
\begin{tabular}{lcccc}
  &  $a_{BX}\ll a_{BB}$  &  $a_{BX}$$\gg$$ a_{BB}$ & $|a_{BX}|$$\gg$$ a_{BB}$ &  $|a_{BX}|\ll a_{BB}$ \\ 
[0.05in]\hline
 $BX^*$$+$$B$~$\rightarrow$~$BB^*$$+$$X$
   &
   ---
   &
   $P_{s^*_0}(\frac{a_{BX}}{a_{BB}})M_{s_0}(\frac{a_{BB}}{r_{0}})a_{BX}$  
   & 
   ---
   & 
   --- \\ 
{\color{White} $BX^*$$+$$B$}~$\rightarrow$~$BB$$+$$X$, $BX$$+$$B$ 
   &
   $P_{s_{0}}(\frac{a_{BX}}{r_{0}})a_{BX}$
   &
   $P_{s^*_0}(\frac{a_{BX}}{a_{BB}})a_{BX}$ 
   & 
   ---
   & 
   --- \\
[0.025in]
$BB^*$$+$$X$~$\rightarrow$~$BX^*$$+$$B$
   &
   $M_{s_{0}}(\frac{a_{BX}}{r_{0}})a_{BX}^2/a_{BB}$
   &
   ---
   & 
   ---
   & 
   --- \\
{\color{White} $BB^*$$+$$X$}~$\rightarrow$~$BB$$+$$X$, $BX$$+$$B$ 
   &
   $a_{BX}^2/a_{BB}$
   &
   $P_{s^*_0}(\frac{a_{BB}}{r_{0}}){a_{BB}}$ 
   & 
   $P_{s^*_0}(\frac{a_{BB}}{r_{0}}){a_{BB}}$ 
   & 
    $P_{s_{0}}(\frac{a_{BX}}{r_{0}})a_{BX}^2/a_{BB}$ \\
[0.025in]
$B$$+$$B$$+$$X$~$\rightarrow$~$BX^*$$+$$B$
   &
   $M_{s_{0}}(\frac{a_{BX}}{r_{0}})a_{BX}^2a_{BB}^2$
   &
   $M_{s^*_0}(\frac{a_{BX}}{a_{BB}})a_{BX}^4$
   & 
   ---
   & 
   --- \\
{\color{White} $B$$+$$B$$+$$X$}~$\rightarrow$~$BB^*$$+$$X$
   &
   $a_{BB}^4$
   &
   $M_{s_0}(\frac{a_{BB}}{r_{0}})a_{BX}^4$
   & 
    $P_{s^*_0}(\frac{a_{BX}}{a_{BB}})M_{s_0}(\frac{a_{BB}}{r_{0}})a_{BX}^4$ 
   & 
   $a_{BB}^4$\\
{\color{White} $B$$+$$B$$+$$X$}~$\rightarrow$~$BB$$+$$X$, $BX$$+$$B$ 
   &
   $a_{BX}^2a_{BB}^2$
   &
   $a_{BX}^4$
   & 
   $P_{s^*_0}(\frac{a_{BX}}{a_{BB}})a_{BX}^4$ 
   & 
   $a_{BX}^2a_{BB}^2$ \\
[0.05in]\hline
  &  $a_{BX}\ll |a_{BB}|$  &  $a_{BX}$$\gg$$ |a_{BB}|$ &  $|a_{BX}|$$\gg$$ |a_{BB}|$ &  $|a_{BX}|\ll |a_{BB}|$ \\ 
[0.05in]\hline
$BX^*$$+$$B$~$\rightarrow$~$BB$$+$$X$, $BX$$+$$B$ 
   &
   $P_{s_{0}}(\frac{a_{BX}}{r_{0}})a_{BX}$
   &
   $P_{s_{0}}^{s^*_0}(\frac{a_{BB}}{r_{0}},\frac{a_{BX}}{a_{BB}})a_{BX}$  
   & 
   ---
   & 
   --- \\
[0.025in]
$B$$+$$B$$+$$X$~$\rightarrow$~$BX^*$$+$$B$
   &
   $M_{s_{0}}(\frac{a_{BX}}{r_{0}})a_{BX}^2a_{BB}^2$
   &
   $M_{s_{0}}^{s^*_0}(\frac{a_{BB}}{r_{0}},\frac{a_{BX}}{a_{BB}})a_{BX}^4$  
   & 
   ---
   & 
   --- \\
{\color{White} $B$$+$$B$$+$$X$}~$\rightarrow$~$BB$$+$$X$, $BX$$+$$B$ 
   &
   $a_{BX}^2a_{BB}^2$
   &
   $a_{BX}^4$ 
   & 
   $P_{s_{0}}^{s^*_0}(\frac{a_{BB}}{r_{0}},\frac{a_{BX}}{a_{BB}})a_{BX}^4$ 
   & 
   $P_{s_{0}}(\frac{a_{BX}}{r_{0}})a_{BX}^2a_{BB}^2$  \\ [0.025in]\hline
%
  &  $a_{XY}$$\gg$$ a_{XZ}$  &  $|a_{XY}|$$\gg$$ a_{XZ}$ 
   &  $a_{XY}$$\gg$$|a_{XZ}|$ &  $|a_{XY}|$$\gg$$|a_{XZ}|$ \\ [0.025in]\hline
$XY^*$$+$$Z$~$\rightarrow$~$XZ^*$$+$$Y$       
   &
   $M_{s_{0}}(\frac{a_{XZ}}{r_{0}})a_{XZ}^2/a_{XY}$
   &
   --- 
   & 
   ---
   & 
   --- \\
{\color{White} $XY^*$$+$$Z$}~$\rightarrow$~$XY$$+$$Z$, $XZ$$+$$Y$, ...
   & 
   $a_{XZ}^2/a_{XY}$
   &
   --- 
   & 
   $P_{s_{0}}(\frac{a_{XZ}}{r_{0}})a_{XZ}^2/a_{XY}$
   & 
   --- \\
$XZ^*$$+$$Y$~$\rightarrow$~$XY$$+$$Z$, $XZ$$+$$Y$, ... 
   & 
   $P_{s_{0}}(\frac{a_{XZ}}{r_{0}})a_{XZ}$
   &
   $P_{s_{0}}(\frac{a_{XZ}}{r_{0}})a_{XZ}$
   &
   ---
   & 
   --- \\ [0.05in]
$X$$+$$Y$$+$$Z$~$\rightarrow$~$XY^*$$+$$Z$ 
   & 
   $a_{XY}^4$
   &
   --- 
   &
   $a_{XY}^4$
   & 
   --- \\
{\color{White}$X$$+$$Y$$+$$Z$}~$\rightarrow$~$XZ^*$$+$$Y$         
   & 
   $M_{s_{0}}(\frac{a_{XZ}}{r_{0}})a_{XZ}^2a_{XY}^2$
   &
   $M_{s_{0}}(\frac{a_{XZ}}{r_{0}})a_{XZ}^2a_{XY}^2$
   &
   --- 
   & 
   --- \\
{\color{White} $X$$+$$Y$$+$$Z$}~$\rightarrow$~$XY$$+$$Z$, $XZ$$+$$Y$, ...
   & 
   $a_{XZ}^2a_{XY}^2$
   & 
   $a_{XZ}^2a_{XY}^2$
   &
   $a_{XZ}^2a_{XY}^2$
   & 
   $P_{s_{0}}(\frac{a_{XZ}}{r_{0}})a_{XZ}^2a_{XY}^2$ \\
\end{tabular}
\end{ruledtabular}
\end{table*}

\begin{figure}[htbp]
\includegraphics[width=3.2in,angle=0,clip=true]{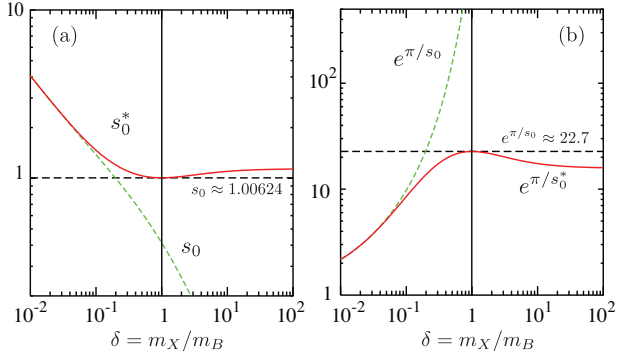}
\caption{(color online). 
Effect of overlapping resonances on (a) the strength of the Efimov potential $s_{0}$ [see Eq.~(\ref{EffPot})]
and (b) the scale factor $e^{\pi/s_0}$ separating Efimov features 
as a function of $a$.
Both panels include the results for large scattering lengths between two 
($s_0$, green dashed line) and three ($s_0^*$, red solid line)
pairs of particles.
}
\label{Fig2}
\end{figure}

Because of their effect on $s_{0}$, overlapping resonances can have a significant impact
on the observation of Efimov physics. In fact, one of 
the major obstacles for observing the 
Efimov effect through ultracold scattering
is that the loss features occur at 
scattering lengths 
separated by the multiplicative 
factor $e^{\pi/s_{0}}$.
Therefore, observing
$n$ features requires the ability to change 
$a$ by roughly a factor of 
$(e^{\pi/s_{0}})^{n}$--- 
$22.7^n$ for identical bosons.
Although a much smaller value for $e^{\pi/s_{0}}$ can be obtained by choosing appropriate mass 
ratios \cite{Us}, this approach severely limits the number of favorable systems.
Near overlapping resonances, however, the situation is drastically improved.
In Fig.~\ref{Fig2} we show $s_{0}$ and 
$e^{\pi/s_{0}}$ for $BBX$ systems where only
$a_{BX}$ is resonant, i.e., for 
an isolated resonance. As can be seen, scaling factors more favorable than 
for identical bosons are obtained only for mass ratios $\delta=m_{X}/m_{B}$ less than approximately 0.2. 
For larger 
$\delta$,
$e^{\pi/s_{0}}$ diverges, making it unlikely that even a single Efimov feature could be observed. 
In contrast, for multiple large scattering lengths,
$s^*_{0}$ and $e^{\pi/s^*_0}$ ---also shown in 
Fig.~\ref{Fig2}
\cite{Efimov}
--- substantially improve
on the identical boson case for {\em all} mass ratios.
Therefore, overlapping resonances dramatically enhance 
the feasibility of observing the Efimov 
effect in a greater variety of systems.

In Tables \ref{TabRates} and II (see Ref.~\cite{EPAPS}), 
we summarize our major results for the collisional rates for $BBX$ and $XYZ$ systems.
In Table~\ref{TabRates}, we assume that only two of the scattering lengths 
are resonant, while in Table~II 
all three scattering lengths are resonant.
For atom-molecule collisions, Table~\ref{TabRates} gives the rate constants for vibrational relaxation, represented generically by
$XY^*+Z\rightarrow XY+Z$, $V_{\rm rel}$, and for reactive scattering, $XY^*+Z\rightarrow XZ^*+Y$, denoted by $\beta$.
For collisions involving three atoms, we give the loss rate for three-body recombination, $K_{3}$,
one example of which is given by $X+Y+Z\rightarrow XY^*+Z$. In Tables \ref{TabRates} and II
the modulation factors $M$ and $P$ are 
\begin{gather}
M_{s_{0}}({\frac{_x}{^y}})\!\propto\!\sin^2[{s_{0}\ln(|\frac{_x}{^y}|)+\Phi}]+\sinh^2\!\eta, \label{M1}\\
P_{s_{0}}({\frac{_x}{^y}})\!\propto\!\frac{\sinh2\eta}{\sin^2[s_{0}\ln(|\frac{x}{y}|)+\Phi]+\sinh^2\!{\eta}},\label{P1}\\
M_{s_{0}}^{s^*_{0}}({\frac{_x}{^y}},{\frac{_z}{^u}})\!\propto\!\sin^2[{s_{0}\ln(|\frac{_x}{^y}|)\!+\!s^*_{0}\ln(|\frac{_z}{^u}|)\!+\!\!\Phi}]\!+\!\sinh^2\!\eta,\label{M2} \\
P_{s_{0}}^{s^*_{0}}({\frac{_x}{^y}},{\frac{_z}{^u}})\!\propto
\!\!\frac{\sinh2\eta}{\sin^2[s_{0}\ln(|\frac{x}{y}|)\!+\!s^*_{0}\ln(|\frac{z}{u}|)\!+\!\Phi]\!+\!\sinh^2\!{\eta}},\label{P2}
\end{gather}
\noindent
and provide the signatures of Efimov physics. 
In these expressions $\Phi$ is an unknown phase 
that determines the positions of the interference minima (in the $M$'s) and the resonant peaks
associated with the creation of an Efimov state ($P$'s), and $\eta$ is a parameter 
associated with the probability to decay 
to deeply bound states. Both $\Phi$ and $\eta$ are non-universal short-range 
three-body parameters that 
must, with the current theoretical limitations, 
be determined 
for each system empirically.

In contrast to isolated resonances, where interference minima only
occur in $K_{3}$ ($a>0$) and resonant enhancements only
occur in $K_{3}$ ($a<0$) and $V_{\rm rel}$ ($a>0$),
minima and peaks appear ubiquitously 
near overlapping resonances and,
in some cases, both can happen in the {\em same} scattering process. 
This result reflects the complexity of
scattering near overlapping resonances and
opens up many alternatives for observing Efimov physics and achieving atomic and molecular 
stability.

We expect the formulas in Tables~\ref{TabRates} and II to apply in the threshold regime \cite{Limits} ---i.e., 
for energies (temperatures) smaller than $1/(2\mu_{BB}a_{BB}^2)$ and $1/(2\mu_{BX}a_{BX}^2)$, where $\mu_{ij}$ is the 
two-body reduced mass--- and when the scattering lengths differ from each other by 
an order of magnitude or more. 
Otherwise, 
the effective potentials will not distinctly display the regions in $R$ discussed above.
In this case, one should assume that 
two similar scattering lengths are equal in magnitude 
for the purpose of using Tables~\ref{TabRates} and II.

\begin{figure}[htbp]
\includegraphics[width=3.2in,angle=0,clip=true]{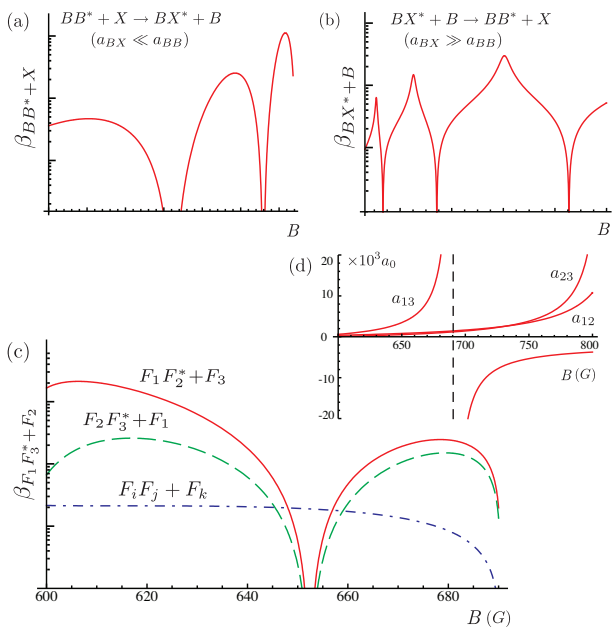}
\caption{(color online) 
The magnetic field dependence of selected processes using the present results.
Reactive scattering showing (a) interference minima and (b) both interference minima and resonant enhancement.
(c) Atom-molecule collision rates for three-spin mixtures of $^6$Li~\cite{3spinsExp} for the region
in (d) where all three scattering lengths 
are large and positive.}
\label{Fig3}
\end{figure}

In Fig.~\ref{Fig3} we single out two interesting cases from Table~\ref{TabRates}. Figure~\ref{Fig3}(a) shows
the interference effects in $BB^*+X\rightarrow BX^*+B$ collisions ($a_{BB}\gg a_{BX}$), and Fig.~\ref{Fig3}(b)
shows the rate for $BX^*+B\rightarrow BB^*+X$ collisions  ($a_{BX}\gg a_{BB}$),
displaying {\em both} interference minima and resonant effects ---the latter is due to the appearance of
Efimov states at the $BX^*+B$ threshold. 
For the examples in Fig.~\ref{Fig3}(a) and (b),
we have chosen an arbitrary 
$a_{BB}$ and $a_{BX}$ dependence on the $B$-field that allows both scattering lengths to vary substantially and to 
display more than one Efimov feature ---note that we also have chosen an arbitrary phase $\Phi$ for the position of the Efimov
features in Eqs.~(\ref{M1})-(\ref{P2}) 
as well as a small 
$\eta$ to make such features more pronounced.
For more realistic systems, decay 
to deeply bound molecules would increase $\eta$, 
possibly limiting the observability of the interference minima, 
In the experiments to date, however, $\eta$ has tended to be small.
Nevertheless, it is interesting to notice that in both cases, if the $a_{BB}$ and $a_{BX}$ can 
be controlled independently
\cite{ScatControl}, one of them can be adjusted to a minimum in $M$ leading to the overall suppression
of the losses as a function of the other scattering length.

In Fig.~\ref{Fig3}(c) we apply our results to the case of the three overlapping resonances
in $^6$Li \cite{3spinsExp} in the range of $B$-field where the three scattering lengths 
$a_{12}$, $a_{13}$, and $a_{23}$, are all large and positive [see Fig~\ref{Fig3}(d)]. 
From our results in Table II (see Ref.~\cite{EPAPS}),
we have found that relaxation of $F_{1}F_{3}^*$ molecules due to collisions with 
$F_{2}$ atoms is suppressed as $a_{13}\rightarrow\infty$. Specifically, we found the 
rates for the processes $F_{1}F_{3}^*+F_{2}\rightarrow F_{1}F_{2}^*+F_{3}$, 
$F_{2}F_{3}^*+F_{1}$ and $F_{i}F_{j}+F_{k}$ (where $F_{i}F_{j}$ is a deeply bound molecule),
to be $M_{s_{0}}(a_{23}/r_{0})a_{12}^2/a_{13}$, $M_{s_{0}}(a_{23}/r_{0})a_{23}^2/a_{13}$, and
$a_{23}^2/a_{13}$, respectively, displaying a $a^{-1}_{13}$ suppression.
These expressions show that $F_{1}F_{3}^*+F_{2}$ collisions can also display interference effects
associated with Efimov physics [see Fig.~\ref{Fig3}(c)]. 
The precise location of the minimum, however, 
will depend on the short-range three-body physics.
Therefore, we can only conclude that 
for this region of $B$-fields the scattering lengths change 
enough to produce a minimum.

In summary, we show that despite the great increase in complexity, the scattering length
dependence of the ultracold three-body rates near overlapping resonances can be understood.
The effect of such overlaps expands the ways in which Efimov physics can be observed
and introduces new effects such as atom-molecule interference minima. We also show that overlapping resonances
reduce the geometric spacing of Efimov features compared to identical bosons for 
essentially all possible 
three-body systems with arbitrary masses.

This work was supported by the National Science Foundation.

\end{document}